# Magnetic Barriers and their $q_{95}$ dependence at DIII-D

F.A. Volpe[1], J. Kessler[2], H. Ali[3], T.E. Evans[4], and A. Punjabi[3]

[1]University of Wisconsin, Madison, WI  53706, USA

[2] Southeast Missouri State University, Cape Girardeau, MO  63701, USA

[3]Hampton University, Hampton, VA  23668, USA

[4]General Atomics, San Diego, CA, USA

**Abstract.** It is well known that externally generated resonant magnetic perturbations (RMPs) can form islands in the plasma edge. In turn, large overlapping islands generate stochastic fields, which are believed to play a role in the avoidance and suppression of edge localized modes (ELMs) at DIII-D. However, large coalescing islands can also generate, in the middle of these stochastic regions, KAM surfaces effectively acting as "barriers" against field-line dispersion and, indirectly, particle diffusion. It was predicted in [H. Ali and A. Punjabi, Plasma Phys. Control. Fusion **49** (2007), 1565-1582] that such magnetic barriers can form in piecewise analytic DIII-D plasma equilibria. In the present work, the formation of magnetic barriers at DIII-D is corroborated by field-line tracing calculations using experimentally constrained EFIT [L. Lao, et al.,, Nucl. Fusion **25**, 1611 (1985)] DIII-D equilibria perturbed to include the vacuum field from the internal coils utilized in the experiments. According to these calculations, the occurrence and location of magnetic barriers depends on the edge safety factor $q_{95}$. It was thus suggested that magnetic barriers might contribute to narrowing the edge stochastic layer and play an indirect role in the RMPs failing to control ELMs for certain values of $q_{95}$. The analysis of DIII-D discharges where $q_{95}$ was varied, however, does not show anti-correlation between barrier formation and ELM suppression.

**1. Introduction**

Chaotic fields [1] and magnetic barriers emerging from chaos [2] recently received a great deal of attention in magnetic confinement research for their capability to, respectively, suppress edge localized modes (ELMs) in tokamaks [3,4] and improve confinement in reversed field pinches [5]. The DIII-D tokamak creates stochastic fields with resonant magnetic perturbations (RMPs), typically of toroidal mode number $n=3$, by means of control coils external or internal to the vessel, called respectively C-coils and I-coils. These stochastic fields reduce the local density gradient, and are experimentally observed to suppress ELMs [1, 3, 4]. This and other means of avoiding or suppressing ELMs are very important for ITER, where ELMs could significantly shorten the lifetime of the divertor and first wall [6]. One of the limitations of RMPs, however, is that experimentally they only suppress ELMs for certain values of the edge safety factor $q_{95}$. Recent evidence suggests that this is connected with a modulation of thermal transport and of the width of the stochastic layer [7], but the fundamental reason of the $q_{95}$ dependence remains still unclear.

Here it is proposed that the modulation of the width of the outermost stochastic layer might be due to the formation and disappearance of magnetic barriers. Kolmogorov-Arnold-Moser (KAM) surfaces [8-11] can form in the middle of chaotic regions. In the context of magnetic confinement, KAM surfaces locally inhibit the dispersion of magnetic field-lines and are therefore also known as magnetic barriers. They were predicted theoretically [12] and may be responsible for the improved particle confinement observed in the TORE-SUPRA [13,14] and TEXTOR [15] experimental results. Parenthetically, to clarify the terminology, barriers "form" during scans of magnetic parameters such as $q_{95}$, everything else remaining fixed. Strictly speaking, however, they are the most resilient laminar surfaces, i.e. the last ones to *disappear* when a perturbation, for example exerted by the I-coils, becomes too strong, everything else remaining fixed.

In the present article we show the first observation of magnetic barriers in the numerical modeling of DIII-D experiments. We also show that, during a $q_{95}$ ramp, the barrier formation is intermittent. Prompted by this observation, we hypothesized that magnetic barriers might contribute to the $q_{95}$ dependence of ELM control. A possible interpretation is as follows. The numerical evidence is that only some values of $q_{95}$ lead to barrier formation. This introduces a laminar surface in the middle of the chaotic layer, thus making the field stochastization "imperfect" or "incomplete". The result is slower transport, i.e. less effective particle pump-out and ELM suppression. It should also be pointed out, however, that the barriers found are very thin, and unlikely to completely inhibit the ELM suppression. They are thus suggested to be a possible concomitant cause, rather than the main one, of the $q_{95}$ dependence of ELM stabilization.

Magnetic barriers were also proposed to play a role in a particular class of internal transport barriers (ITBs), not related with reversed shear ITBs, and apparently forming at rational surfaces [16, 17]: these rational $q$ transport barriers were attributed in [18, 19] to magnetic barriers forming at irrational values of $q$ in the immediate vicinity of those rational surfaces.

Finally, it was proposed that magnetic barriers might also play a role in locked mode disruptions [20].

Before studying the consequences of magnetic barriers for ELMs, transport barriers and locked modes, however, it is important to assess whether magnetic barriers are actually feasible in realistic magnetic configurations, in the first instance. In the present article we positively answer this question for DIII-D, by means of two computer codes.

The paper is organized as follows. The theory of magnetic barriers is briefly reviewed in Sec.2. In Sec.3, a code previously applied to the prediction of $n=1$ barriers at DIII-D [21] is modified to show that under the same idealized conditions, barriers can form at DIII-D as a result of overlapping $n=3$ islands. Note that this is the same toroidal number successfully used in ELM control experiments at DIII-D. Criteria for the automatic detection of islands, of laminar surfaces and chaotic regions are formulated and tested against these idealized data. In Sec.4 these criteria are applied to more realistic field-line tracings in DIII-D equilibria perturbed by the I-coil vacuum field. It is shown that laminar surfaces can only form in the middle of stochastic regions for special values of $q_{95}$ which, however, do not correlate well with the values at which RMPs fail to stabilize ELMs. Finally, the appendix provides details on a numerical method used to discriminate between laminar surfaces, magnetic islands and stochastic regions in a given Poincaré plot.

## 2. Background: "building" magnetic barriers in tokamaks

Magnetic barriers were observed numerically in [21] in special toroidal configurations obtained by adding a *non-linear* control term to a piecewise analytic tokamak equilibrium featuring overlapping islands. The initial equilibrium is described, in magnetic co-ordinates $\psi$, $\theta$ and $\phi$ (toroidal flux, poloidal angle and toroidal angle, respectively), by the poloidal flux

$$\chi(\psi,\theta,\phi) = \chi_0(\psi) + \varepsilon\, \chi_1(\psi,\theta,\phi) \tag{1}$$

i.e. by the sum of an unperturbed poloidal flux $\chi_0(\psi)$ corresponding to the unperturbed equilibrium (laminar flux surfaces) and a first-order perturbation $\chi_1$. The latter is given by the *linear* superposition of the poloidal fluxes associated with individual islands of poloidal number *m* and toroidal number *n*,

$$\chi_1(\psi,\theta,\phi) = \sum_{m,n} \chi_{mn}(\psi)\cos(m\theta - n\phi + \delta_{mn}), \qquad (2)$$

where $\delta_{mn}$ are phases. The coefficient $\varepsilon$ in Eq.1 is dimensionless and quantifies the strength of the magnetic perturbation. In our case, this is proportional to the intensity of the I-coil currents. Fluxes $\chi_{mn}(\psi)\cos(m\theta - n\phi + \delta_{mn})$ in Eq.2 are the island poloidal fluxes for $\varepsilon=1$.

As it is well known, for sufficiently large $\varepsilon$, the flux $\chi$ in Eq.1 develops chaotic structures. The idea of [21], based on control theory and Hamiltonian mechanics [22-25], is that the addition of a special "control term" can make the poloidal flux $\chi$ more regular, less chaotic than in the original system. As a result, laminar surfaces become more resilient, i.e. they break up into chaos at higher values of $\varepsilon$. Such resilient laminar surfaces (or KAM surfaces, in the language of Hamiltonian systems), in particular those located in the middle of chaotic regions, will hereafter be called magnetic barriers. The reason is that, being laminar surfaces in the middle of chaos, they act as local "barriers" against the dispersion of field-lines.

At the lowest significant order, the control term to add to Eq.1 to tame chaos and lead to the formation of an invariant torus at $\psi_b$ is [21]

$$\varepsilon^2 \chi_2(\theta,\phi) = -\varepsilon^2 \frac{\iota'(\psi_b)}{2}\left[\sum_{m,n}\frac{m\chi_{mn}(\psi_b)}{m\iota(\psi_b)-n}\cos(m\theta - n\phi + \delta_{mn})\right]^2 \qquad (3)$$

where $\iota(\psi_b)$ is the rotational transform, $\iota = 2\pi/q$, evaluated at $\psi_b$ and $\iota'(\psi_b)$ is its derivative with respect to $\psi$. When this second order term is added to the perturbed Hamiltonian (poloidal flux), it forms an invariant torus at $\psi=\psi_b$, and this invariant manifold or magnetic barrier continues to exist for up to some maximum value of the magnetic perturbation, and inhibits the dispersion of magnetic field lines in its vicinity. This is because field lines cannot cross an invariant manifold. The position of the rational surfaces $\psi_{mn}$'s and the relative positions of the barrier at $\psi_b$ and the rational surfaces depend upon the safety factor profile $q(\psi)$. This means that $q_{95}$ can have impact on the formation and resilience of the magnetic barriers. In this paper we investigate this issue in the case of the DIII-D tokamak.

The primary goal of the present paper is to determine whether magnetic barriers are feasible or have been already, serendipitously obtained at DIII-D. After giving a positive answer to this question, we study how robust these barriers are to changes of the magnetic configurations (I-coil currents and $q_{95}$), finding that they alternatively form and disappear during a $q_{95}$ ramp. Some correlation is found with ELM control by RMPs, which is also intermittent. It is suggested that the formation of barriers temporarily reduces the width of the edge stochastic layer. This would reconcile our findings with the observation [7] that ELMs are suppressed or not, depending on whether the stochastic edge is wide enough or is too narrow.

Note that Eq.3 is quite general with respect to the island location $\psi_b$. Some locations, however, are more robust than others [21,26]. Eventually, in the presence of a strong enough perturbation, all laminar surfaces stochastize. The last one to do so is called the most resilient or "most noble" barrier. From number theory [27], this surface coincides with the "most irrational" surface, where $q$ takes the continued fraction value

$$q = a_0 + \cfrac{1}{a_1 + \cfrac{1}{a_2 + \cdots}},  \qquad (4)$$

where $a_0=a_1=a_2=\ldots=a_n=1$. This value of $q$ is the most irrational number called the golden ratio, $\varphi=(1+\sqrt{5})/2=1.618...$

Figs.6-9 of [21] provided examples of barriers breaking in cantori and small islands when the perturbation is too strong or the barrier location is non-optimal.

## 3. Detection of magnetic barriers in idealized geometries

This Section presents numerical diagnostics of islands, stochastic regions and laminar surfaces, hereby including magnetic barriers. First of all, $\psi$-$\theta$ maps are generated, for the safety factor q($\psi$) of the DIII-D tokamak [21], and magnetic barriers are "built" as described in Sec.2. Subsequently, "diagnostics" are developed and tested on these idealized data. In the following Sections the diagnostics will be applied to puncture plots obtained with the TRIP3D field-line tracer code [28] in the DIII-D configurations.

Fig.1a shows an example of chaos generated by overlapping islands of toroidal number $n=3$, as used in ELM control experiments. Specifically, $m/n=8/3$ and $9/3$ islands are considered, and summed over in Eq. 2, using $\varepsilon = 5.25 \times 10^{-4}$. For simplicity, the phases $\delta_{mn}=0$ for both modes. This means that these two resonant modes are locked. One can look upon this magnetic perturbation either as locked in the laboratory frame, or as an instantaneous snapshot of a rotating magnetohydrodynamics mode in the lab frame. The figure shows the field-line puncture plot in a given poloidal section at a fixed toroidal location, $\phi = 0$. Ten field-lines of the same initial $\theta$ and equally spaced initial $\psi$ are followed for 1000 toroidal turns. If one adds to Eq.1 the second-order term of Eq. 3, with the same $\iota$ profile as in [21], and with $\psi_b=0.6971$, the plot modifies as in Fig.1b. The chaos of the system is greatly reduced, and a magnetic transport barrier forms at the prescribed $\psi_b$, as expected. Similar graphs were published in [21] for $n=1$.

While islands, chaos and a barrier are clearly visible in Figs. 1a-b, they are more difficult to recognize in realistic field-line tracing maps. Three methods that partly automate the recognition and differentiation of these three types of "structures" are presented and discussed in the remainder of the Section and are illustrated in Figs. 2-4.

*3.1. Radial Width*

The first method consists in analyzing the interpolated $\psi$, $\bar{\psi}$, and the $\psi$-width, $w$, of each structure. This analysis is performed in a single Poincaré plot, i.e. at fixed $\phi$. The quantities $\bar{\psi}$ and $w$, however, are functions of $\theta$. Here $\bar{\psi}$ denotes an interpolated function $\bar{\psi}(\theta)$ which at most is single-valued (Figs. 2a,c) while $w(\theta)$ denotes the local radial width of the structure, in units of $\psi$, at a certain $\theta$ (Figs. 2b,d). Neither $\bar{\psi}(\theta)$ nor $w(\theta)$ are necessarily defined at all $\theta$. For example if and only if the structure is an island, they exhibit periodic poloidal gaps, which can thus be considered a simple and unique indicator of islands (Figs. 2a-b).

By contrast, both laminar surfaces and stochastic regions (if observed for a sufficient number of toroidal turns) cover all possible values of $\theta$, i.e. they don't present any poloidal gap $\Delta\theta$. However, there are other characteristics that permit us to differentiate one from the other. The most obvious is that, unlike *laminar* surfaces, stochastic layers have finite $w(\theta)$ (Fig. 2d). Note, however, that the width $w(\theta)$ of these complicated 3D objects is not the same in every $\theta$. It is therefore convenient to introduce the *maximum* and *average* width of a structure, $w_{max}$ and $\bar{w}$. Here the *maximum* width $w_{max}$ indicates the maximum width that a structure has in some $\theta$. The *average* width $\bar{w}$, instead, is the poloidally averaged radial width, in units of $\psi$.

$w_{max}$ and $\bar{w}$ of the structures in Fig. 1b are plotted in Fig. 3a, on the horizontal axis, as a function of their initial $\psi$ location in the field-line calculations, $\psi_0$, on the vertical axis. Both $w_{max}$ and $\bar{w}$ *nearly* vanish at $\psi_0$=0.693 in Fig.3a (due to numerical reasons they never vanish exactly, even for perfect laminar surfaces). This value is in good agreement with the position, $\psi_b$=0.697, of the magnetic barrier in Fig. 1b. In Fig.3a, however, $\bar{w}$ also reaches very small values at locations ($\psi_0$=0.681, 0.709 and, possibly, 0.674) where there are no barriers. It is concluded that $w_{max}$ yields a more selective criterion. In brief, laminar surfaces have zero width (within numerical uncertainty), hence a search of zeros (or minima) of the maximum width $w_{max}$ identifies candidate barriers.

*3.2 Field line Dispersion*

An additional criterion relies on the average field line dispersion, which is sometime referred to as a field line diffusion coefficient [21],

$$\langle D_F \rangle = \frac{1}{500} \sum_{i=1}^{500} \frac{(\psi_{N,i}-\psi_{0,i})^2}{4\pi N}. \tag{5}$$

This is calculated by tracing, for *N* toroidal turns, 500 field lines of fixed initial $\psi_0$ but different initial $\theta$ and then averaging $\frac{(\psi_{N,i}-\psi_{0,i})^2}{4\pi N}$ over those 500 field lines. Here $\psi_{N,i} - \psi_{0,i}$ is the variation of $\psi$ after *N* toroidal periods. *N*=500 is used here. Fig.3b shows <$D_F$> as a function of $\psi_0$. A clear minimum in the dispersion can be recognized in Fig.3b at the same $\psi_0$ where zero width was recorded in Fig. 3a, $\psi_0$=0.693. With good approximation, this coincides with the expected barrier location, $\psi_b$=0.681. The small discrepancy is attributed to the use of the *unperturbed* flux to localize the region of minimum width and minimum dispersion (Fig. 3) in the *perturbed* equilibrium of Fig. 1, where the perturbed surfaces do not have the same $\psi$ everywhere (at all $\theta$) and in particular not the same initial field-line tracing value $\psi_0$. Here, it is noted that the change in the field line topology produced by the vacuum RMP fields is not a time-dependent process, so we choose to refer to this change in the field-line topology as a *dispersion* of the field lines rather than a diffusion of the field lines, which implies a time-dependent process.

*3.3 Poloidal Spectrum*

Yet another criterion for distinguishing islands, chaos and, indirectly, laminar surfaces is offered by the poloidal spectrum (Fig.4): for each of the 10 field lines traced in Fig.1, the interpolated function $\bar{\psi}(\theta)$ and the width $w(\theta)$ (Fig. 2) are expanded in poloidal harmonics $\psi_m \sin(m\theta)$ and $w_m \sin(m\theta)$. The amplitudes $\psi_m$ and $w_m$ of the poloidal components are plotted in the form of contours, as functions of the surface label $\psi_0$ and poloidal number $m$ (Fig. 4). "Surfaces" at $\psi_0 \lesssim 0.68$ and $\psi_0 \gtrsim 0.70$ correspond respectively to the 8/3 and 9/3 islands, and to their immediate vicinity. Indeed these surfaces exhibit peaks at $m=8$ and $m=9$ in both contours. They also present peaks at the harmonics and sub-harmonics 4, 12, 16, etc. for $m=8$ and 3, 6, 12, 15, etc. for $m=9$. Surfaces at $0.68 \lesssim \psi_0 \lesssim 0.70$ have mostly broken in chaos (except for the barrier). The poloidal spectra are broader, reflecting the fact that they don't have the ordered periodic nature of islands. Still, as they are bordered by the 8/3 and the 9/3 island, both $\bar{\psi}(\theta)$ and $w(\theta)$ experience some modulation at $m=8$-9, captured by peaks in their poloidal spectra (Fig. 4). The biggest peak, however, is at $m=0$, representing a poloidally flat radial offset of the interpolation $\psi(\theta)$ (Fig. 4a) or a finite poloidal average of the structure width (Fig. 4b). Fig.4a does not allow us to differentiate magnetic barriers from chaos. However, the absence of an $m=0$ peak in Fig. 4b can be used to identify the surfaces at $\psi_0 = 0.681$, 0.693 and 0.709 as *candidate* magnetic barriers. This is due to the fact that the $m=0$ component of the poloidal expansion of $w(\theta)$ coincides with its average, $\bar{w}$, and $\bar{w} = 0$ is a necessary, although not sufficient, condition for magnetic barriers (see Fig. 3a).

Finally, $\psi$ is a periodic function of the toroidal angle (not shown) in the case of islands and laminar surfaces but not, evidently, for chaos.

The above criteria are summarized in Table 1. These criteria were used to automatically recognize islands, stochastic regions and laminar surfaces in DIII-D Poincaré plots. Details on this numerical procedure are presented in the Appendix.

*Table 1: Criteria for the recognition of magnetic islands, stochastic regions and laminar surfaces, hereby including magnetic barriers*

|  | **Island** | **Stochastic Region** | **Laminar Surface** |
|---|---|---|---|
| **Max Radial Width** $w_{max}$ | Finite | Finite | =0 |
| **Max Poloidal Gap** $(\Delta\theta)_{max}$ | Large | Small | =0 |
| **Field-line Dispersion** $\langle D_F \rangle$ | Medium | Large | Small |
| **Poloidal Spectrum of** $\bar{\psi}$ | Peaked at resonant $m$, its harmonics and sub-harmonics | Broad, peaked at $m=0$ | Broad, peaked at $m=0$ |
| **Poloidal Spectrum of** $w$ | Peaked at resonant $m$, its harmonics and sub-harmonics | Broad, peaked at $m=0$ | =0 |
| **Toroidal Dependence** | Periodic | Non-periodic | Periodic |

## 4. Detection of magnetic barriers in the numerical modeling of DIII-D experiments

Here $n$=1 (Subsec. 4.1) and $n$=3 (Subsec. 4.2) I-coil vacuum fields are superimposed on the unperturbed EFIT [29] equilibria for DIII-D discharges. The field lines are traced by means of the TRIP3D code [28], and the criteria of Table 1 are applied, confirming the prediction of barrier formation between 3/1 and 4/1 islands [21] and showing that barriers can form during ELM control experiments by $n$=3 RMPs.

Barriers are also studied during an I-coil current ramp (Subsec.4.3) and two $q_{95}$ ramps (Subsec.4.4 and 4.5).

*4.1 Odd n=1 magnetic barrier*

To begin with, let us consider the DIII-D equilibrium from shot 115467 at 3000 ms. To test the prediction that barriers can form at DIII-D between 3/1 and 4/1 islands [21], we apply an RMP of toroidal periodicity $n$=1 in which the upper and lower row of I-coils are out of phase by 180$^o$. I-coil currents of 1.45 kA result in large 3/1 and 4/1 islands at $\psi$=0.86-0.89 and $\psi$>0.95, respectively (Fig.5). The perturbation considered is large. For comparison, typical DIII-D error fields can be corrected by I-coil currents of less than 1 kA. As a result of the large perturbation, the islands in Fig.5 are large, and the region in between is highly stochastic. At the same time, however, a magnetic barrier is clearly visible at $\psi\approx$0.91, approximately half way between the 3/1 and 4/1 islands, as predicted [21]. Additionally, the barrier appears to be reinforced by higher $m$ island chains (13/4 and 10/3), which was also predicted, as a result of higher order corrections to Eqs.1-3 [21]. The barrier is immediately recognizable by visual inspection of Fig.5, and is confirmed by zeros in width (Fig.6a) and diffusion coefficient (Fig.6b) as well as by the low and broad poloidal spectrum (Fig.7b).

Note that the Poincaré plot in Fig.5 was obtained by numerically adding an $n=1$ RMP to an equilibrium (shot 115467 at t=3000ms) calculated in the absence of field errors. Experimentally, a similar result can be obtained by applying (or leaving uncorrected) an $n=1$ error field of appropriate amplitude (corresponding to 1.45kA I-coil current) to a discharge similar to 115467.

*4.2 Odd, even and single row n=3 magnetic barrier*

I-coil perturbations of $n=3$ were superimposed on the unperturbed equilibrium for shot 115467 at 3000 ms. Three $n=3$ configurations were considered: up-down symmetric (even parity), asymmetric (odd parity) and utilizing a single row of I-coils (the upper one).

I-coil perturbations of the same current intensity (3 kA) result in a highly stochastic field in the even parity case (Fig. 8a) and in relatively small, non-overlapping islands and various laminar surfaces in the odd parity case (Fig. 8c). An intermediate behavior is obtained when a single row of I-coils is energized (Fig. 8b). In other words, even parity is more effective at stochastizing the field, in agreement with the fact that, experimentally, it is more effective at suppressing ELMs.

*4.3 Barrier destruction during I-coil ramp-up*

Figs.9a-b illustrate how sufficiently intense RMPs perturb the equilibrium flux surfaces (Fig.9a) by opening up magnetic islands (Fig.9b). An $n=1$ case is considered here. Higher I-coil currents $I_{\text{I-coil}}$ lead to larger islands that begin to overlap. This marks the onset of field stochastization (Fig.9c). At higher and higher $I_{\text{I-coil}}$, more and more laminar surfaces are replaced by islands or stochastic layers (Fig.9d). Eventually, all laminar surfaces are destroyed. The last laminar surface

to break, at $\psi \cong 0.905$ in Fig. 9e, is the most noble magnetic barrier. The data were obtained by TRIP3D, by superimposing $n=1$ I-coil vacuum field to the EFIT equilibrium for shot 115467 at time $t=3000$ms, i.e. at a fixed $q_{95}$.

The increase in stochasticity as $I_\text{I-coil}$ increases is documented by the growing field-line dispersion in Fig.10. After the onset of stochastization (marked) $<D_F>$ increases more rapidly with $I_\text{I-coil}$ as more and more laminar surfaces are destroyed. As soon as the last barrier is destroyed, $<D_F>$ continues to grow linearly with $I_\text{I-coil}$. Obviously, a plasma edge featuring laminar surfaces, hereby including barriers ($I_\text{I-coil}$ <1575A, in the example considered) is less stochastic (has lower $<D_F>$) than a plasma edge without laminar surfaces ($I_\text{I-coil}$ >1575A). In other words, Fig.10 quantifies how the presence of barriers makes the edge less stochastic.

*4.4 Intermittent barrier formation during $q_{95}$ ramps*

Analysis of the $n=3$, even parity DIII-D discharge 132741 demonstrated a correlation between ELM suppression and a broadening of the edge stochastic layer while $q_{95}$ was ramped down from 4 at 2400 ms to 3.2 at 5000 ms [7].

Here the same discharge is studied from the point of view of magnetic barriers, using the criteria outlined in Sec.2. Poincaré plots were generated by means of the TRIP3D code for different time slices, at intervals of 50 ms, corresponding to small $q_{95}$ variations, $\Delta q_{95}=0.015$. Such a scan is fine enough to resolve intervals of successful ELM suppression ($q_{95}=3.18$-3.26, 3.33-3.65 and 3.77-3.91 [7]). To test whether there is any correlation with magnetic barriers, radial profiles of structure width similar to Fig. 3 and 6 and poloidal spectra in the way of Fig. 4 and 7 were generated for each puncture plot. Furthermore, the maximum poloidal gap $\Delta\theta$ was also monitored for each structure. Magnetic islands tend to exhibit large poloidal gaps, as shown previously in Fig.2a. This characteristic allows to discriminate radially narrow islands from laminar surfaces. Some structures, such as large islands and large stochastic layers are clearly recognizable by visual inspection of the Poincaré plot (Fig. 11a), but for other structures it is necessary to examine the widths (Fig.11b), field-line dispersion (Fig.11c) and poloidal gaps profiles (Fig.11d). Then, with the aid of the criteria in Table 1, it can be determined whether the structure at a certain $\Psi_0$ and at a given time ($t$=2550ms in Fig. 11), is an island, a stochastic layer or a laminar surface. The automatic analysis of plots similar to Figs. 11b-d was repeated for different times and summarized in Figs.12a-c in the form of contour plots. Figs.12a-b document the presence of radially localized (low $\Delta\Psi$), low-dispersion (low $<D_F>$) structures in the middle of the high-dispersion stochastic edge of the DIII-D plasma. These structures are moved towards the very edge ($\Psi$=1) by the evolution of the $q$ profile, as $q_{95}$ is decreased (Fig.12e). Not all of these narrow structures are laminar surfaces, however: several of them exhibit large poloidal gaps (Fig.12c) and are thus identified as islands.

Depending on which criteria in Table 1 are fulfilled at each $\Psi_0$ and $t$ in Figs. 12a-c, we can categorize and color-code the corresponding structure, and synthesize the information in a fourth contour plot, Fig. 13a. This allows to distinguish and spatially and temporally resolve laminar surfaces, Cantori and island O-points (respectively characterized by no poloidal gaps, small ones, or large ones) as well as stochastic regions or large overlapping islands. Here Cantori or Cantor sets [19] are chains of very thin, elongated islands, consistent with barriers which just broke as a result of a slightly above-marginal magnetic perturbation, as predicted in Ref. [21]. The Cantori found (Figs. 14a-b) are remarkably similar to theoretical predictions (Figs. 14c-d) and, even though they are not perfect barriers, they are very effective anyway at preventing or limiting the dispersion of field lines, for example from the inner region marked in red to the outer region marked in blue in Fig. 14.

Magnetic barriers intermittently breaking in Cantori are observed in Fig.13a at $t >$ 3250ms, corresponding to $q_{95} < 3.7$. With time, as a result of the reduced $q_{95}$, the barriers drift towards outer radii. At $t$=4750ms a barrier forms at a much outer $\psi_0$. If one assumes the thickness of the edge stochastic layer $d$ to be determined by the *laminar* surface bounding such a stochastic layer, then Fig.13 and the time series of Poincaré plots in Fig.15 would suggest a gradual thinning of the edge stochastic layer, followed by a more abrupt change at $t$=4750ms. While the latter seems to correlate with failed ELMs suppression at $t > $ 4750ms in Fig. 13c, the gradual drift at $t$=3250-4750ms does not correlate with alternate failures and success of controlling ELMs in that interval (Fig. 13c). Therefore magnetic barriers do not seem to explain the $q_{95}$ dependence of ELM control, at least according to the present TRIP3D calculations which were performed in vacuum, i.e. did not include the plasma response.

A different discharge where $q_{95}$ was ramped up rather than down (Fig.17b) was also analyzed (Figs. 17-18). Again, magnetic barriers can form and do depend on $q_{95}$ (Fig.17a). Poincaré plots show persistent laminar surfaces at $\Psi \simeq 0.8$-$0.82$, but intermittent ones at $\Psi \simeq 0.86$ (Fig.18). This suggests a modulation of $d$ with $q_{95}$ which however, unfortunately, does not anti-correlate with ELM suppression in Fig.17c.

## 5. Summary and Conclusions

In summary the feasibility, in the DIII-D tokamak, of magnetic barriers predicted in [21] under idealized conditions is confirmed here using experimentally constrained EFIT [29] equilibria and vacuum field perturbations exerted by the internal I-coils, taking into account their actual geometry. Barriers are observed in a numerical model of the DIII-D edge, stochastized by an $n$=1 field comparable with the uncorrected machine error field. Perfect and imperfect barriers, broken into cantori as a result of too strong magnetic perturbations are observed as a result of $n=3$ even parity perturbations, as used for ELM control. The formation of barriers, as well as their location exhibit a dependence on $q_{95}$, but little or no correlation is found with the failure of the RMPs at suppressing ELMs. The plasma response is currently not included in the model, but it is expected that its inclusion will confirm the feasibility of magnetic barriers. The reason is that the plasma response has been hypothesized to shield the RMP fields deep in the plasma, near the $q=2$ surface, and may amplify the field at outer locations ($\rho > 0.9$). Understanding the response of the plasma to small (~ $10^{-4}$ $B_T$) 3D magnetic perturbations fields involves a variety of complex physics issues and is currently an area of active research.

Note that, while potentially detrimental for ELM control, magnetic barriers can be beneficial for core confinement.

**Acknowledgements**

Work supported in part by the US DOE under a National Undergraduate Fusion Fellowship and DE-FC02-04ER54698, additional financial support from UW-Madison. JK acknowledges fruitful discussions with J. Burby (Cornell Univ.) and A. Avril (Univ. of Washington) and the hospitality of General Atomics and the University of Wisconsin, Madison.

**Appendix – Numerical recognition of Barriers, Cantori, Islands and Chaos**

It was shown in Sec.3 and Figs.2-7 that different structures (islands, stochastic regions, laminar surfaces, hereby including barriers) have different characteristics (radial widths, poloidal spectra, toroidal dependences). These differences were summarized in Table 1, and were used in Sec. 4 to automatically recognize these structures in DIII-D Poincaré plots, computed for example with TRIP3D. The automatic procedure works as follows. A given Poincaré plot is treated as $N_s$ sets of ($\psi,\theta$) points. Each set of point corresponds to a single field-line, which was traced for 500 toroidal turns in TRIP3D: the ($\psi,\theta$) points are the intersections of that field-line with the poloidal cross section examined. The Poincaré plots examined were very densely populated, with the field-line initial conditions spaced by as little as $\Delta\psi_0$=0.002. For each set of points, or "structure", the maximum width $w_{max}$ was calculated as in Sec.3.1. As discussed, stochastic regions and large islands are characterized by a large $w_{max}$. There is some degree of arbitrariness in how large this should be, but inspection of Figs.3, 6 and 11 suggested $w_{max}$>0.01 to be a reasonable criterion. Indeed, this was successfully tested on several Poincaré plots and ultimately adopted for the automatic recognition of stochastic regions and large islands.

The remaining structures, with $w_{max}\leq0.01$, were categorized as laminar, cantori or island O-points on the basis of their maximum poloidal gap $(\Delta\theta)_{max}$. Points ($\psi,\theta$) belonging to a laminar surface are expected to have the most uniform distribution in $\theta$. In the large aspect ratio limit, gaps between consecutive points are expected to be all of the order of $\Delta\theta\approx0.72^o$ (corresponding to 500 points equally spaced over 360$^o$). However, due to finite aspect ratio, the high-field side is more densely populated with field-lines, approximately twice as much as the low-field side (at the outer minor radii of interest here). Additionally, neighboring islands and the divertor make the poloidal distribution of points on a laminar surface even less uniform. In conclusion, $(\Delta\theta)_{max}<10^o$ was found to be a better criterion to identify laminar surfaces, including magnetic barriers. Cantori are chains of elongated islands separated by small gaps, but larger than for laminar surfaces: the criterion adopted was $10^o \leq (\Delta\theta)_{max} < 20^o$. Finally, large gaps $(\Delta\theta)_{max} > 20^o$ are a definite indicator of island O-points. Note that in the continuous limit, i.e. of infinite toroidal turns, laminar surfaces are single-valued in $\theta$, whereas cantori and islands are double-valued.

Other criteria successfully tested in Sec.3 and summarized in Table 1 include examination of the poloidal spectra of $\bar{\psi}$ and $w_{max}$, and toroidal dependence, but these were found to be computationally expensive for the analysis of many, densely populated Poincaré plots as needed for the analysis of an entire discharge. For this reason, they were not used for the analysis presented in Figs.12, 13, 16 and 17.

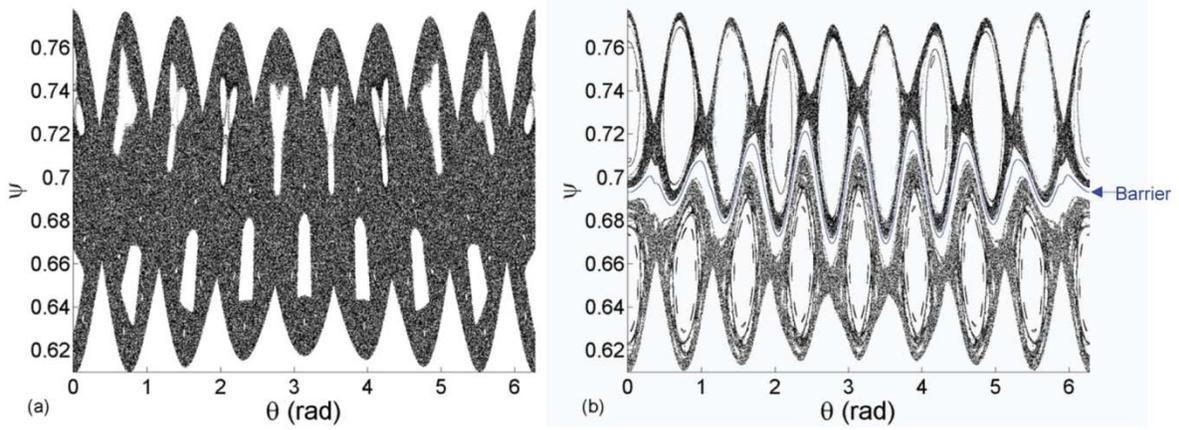

Figure 1 (a) Poincaré plot of stochastic surfaces between 8/3 and 9/3 islands for $\delta = 5.25 \times 10^{-4}$. (b) Poincaré plot of stochastic surfaces between 8/3 and 9/3 islands for $\delta = 5.25 \times 10^{-4}$ with second order control term.

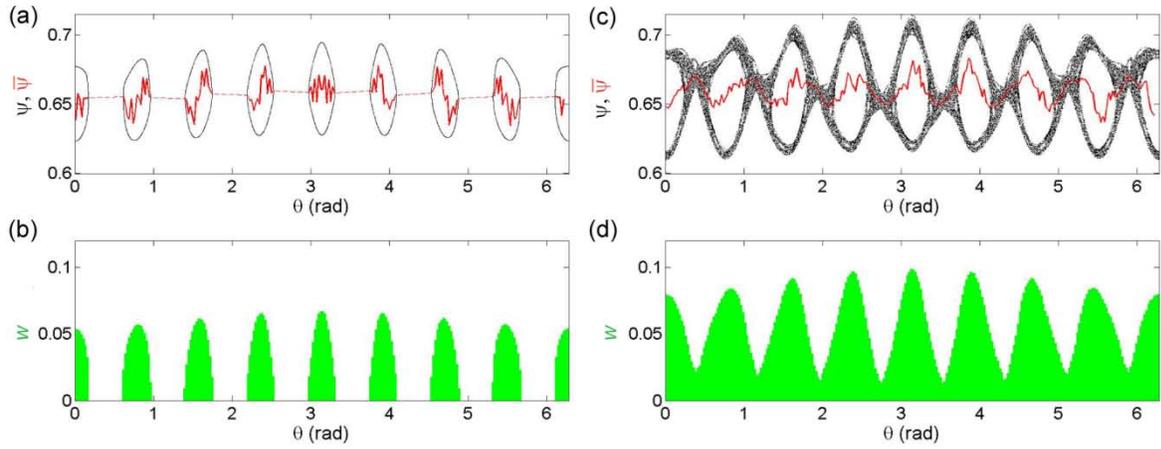

Figure 2 (a) Magnetic surface corresponding to 8/3 island with interpolated function $\bar{\psi}$ overlaid in red. (b) The width *w* of the 8/3 island is finite and exhibits poloidal gaps. (c) Stochastic structure surrounding 8/3 island with interpolated function $\bar{\psi}$ overlaid in red. (d) Width *w* of stochastic structure is finite at all poloidal angles $\theta$.

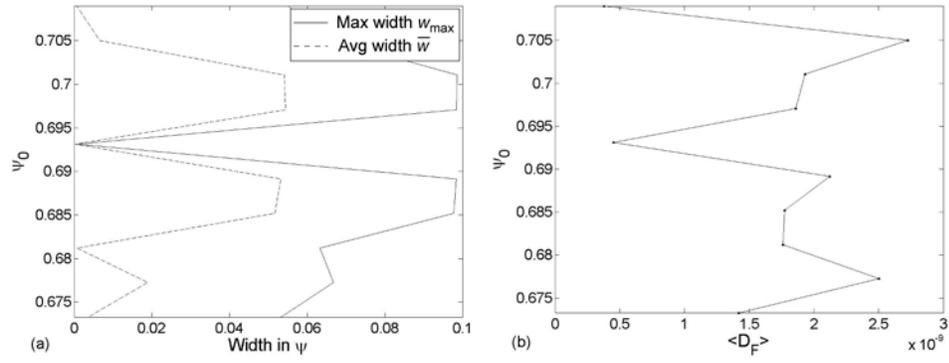

Figure 3 Radial profiles of (a) maximum and poloidally averaged width and (b) field-line dispersion coefficient $<D_F>$ for the structures of Fig.1. Profiles exhibit minima at $\psi_0=0.693$, in correspondence of the magnetic barrier.

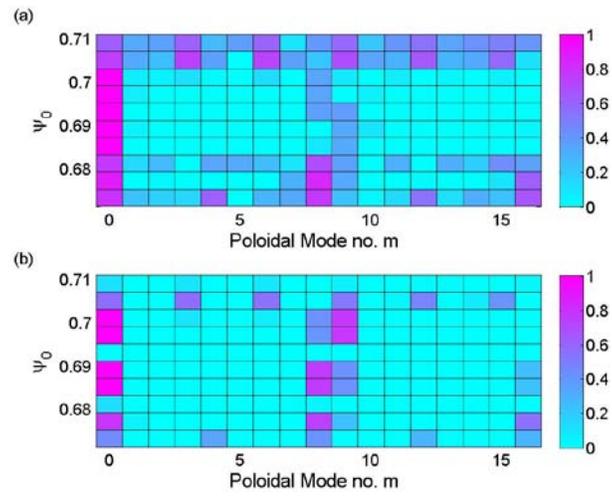

Figure 4 Poloidal spectra of (a) $\bar{\psi}(\theta)$ and (b) $w(\theta)$ for the structures of Fig.1. Notice peaks at $m=3$, 8, 9 and their harmonics, due to 8/3 and 9/3 islands. Notice also low, broad $w(\theta)$ spectrum in correspondence of barrier ($\psi_0=0.693$), but also in other locations.

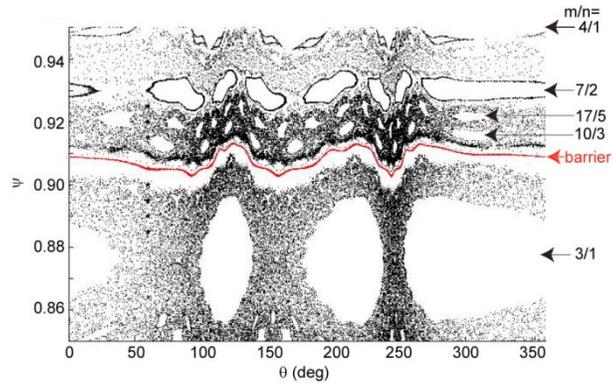

Figure 5 Poincaré plot of 10 initial values for an *n*=1 odd parity, 1450A I-coil perturbation after 10 000 toroidal turns with robust barrier shown in red.

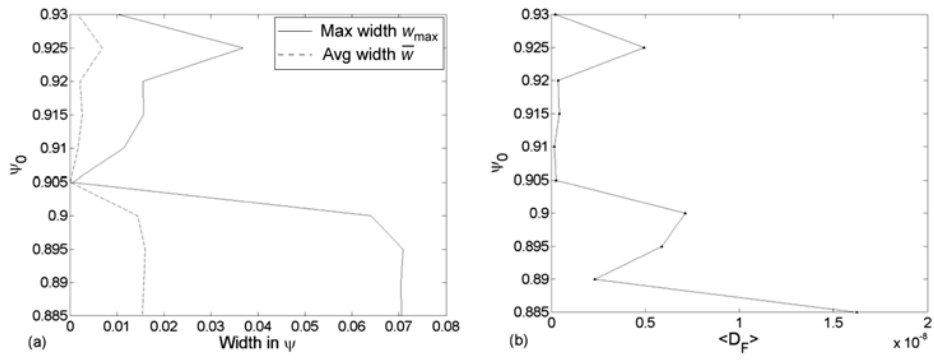

Figure 6 Like Fig.3, but for the structures of Fig.5.

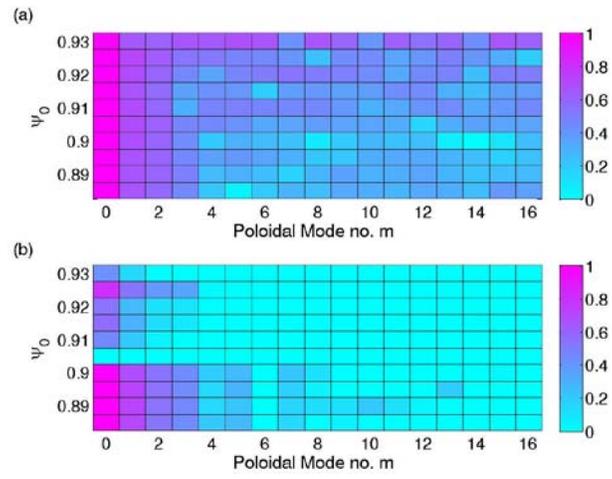

Figure 7 Like Fig. 4, but for the structures of Fig. 5.

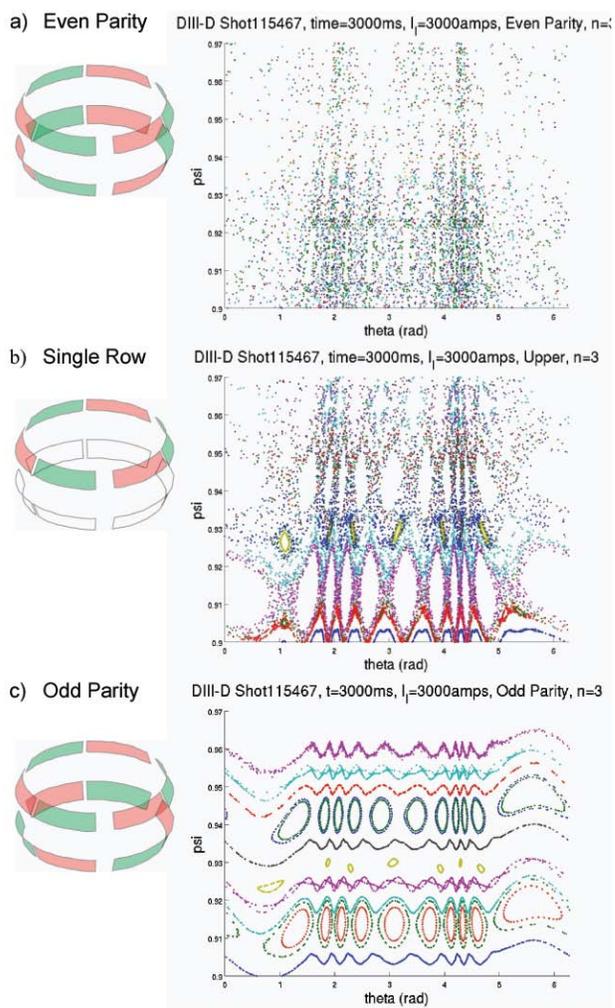

Figure 8 Poincaré plots for (a) even parity, (b) single row, and (c) odd parity *n*=3 I-coil perturbations.

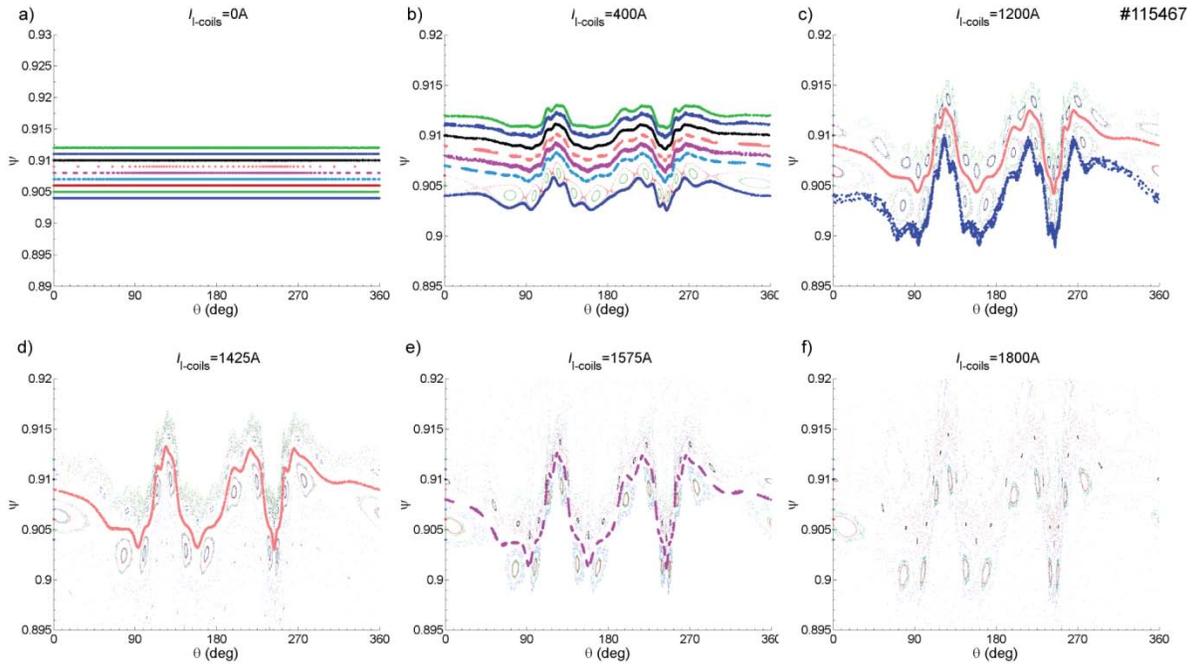

Fig.9 RMPs of toroidal mode number $n$=1 and increasing strength (increasing I-coil current) progressively stochastize and destroy laminar surfaces in discharge 115467. The most resilient laminar surface is, by definition, a magnetic barrier. Stars represent the initial conditions of the field-lines traced. (a) Laminar unperturbed flux surfaces. (b) Magnetic islands form for small perturbations. (c) Islands begin to overlap, leading to field stochastization. (d) Fewer and fewer laminar surfaces survive to stronger and stronger perturbations. (e) At I-coil currents of 1575A, the last laminar surface breaks in a cantorus. (f) For even bigger perturbations, the field is completely stochastic.

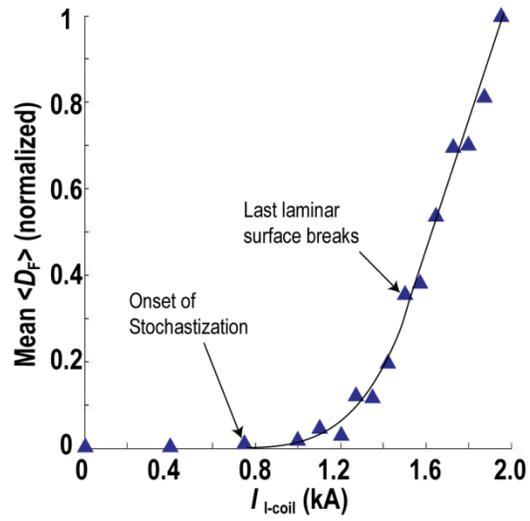

Fig.10 Magnetic field line diffusivity associated with the *n*=1 Poincaré plots of Fig.9, as a function of the I-coil current $I_{\text{I-coil}}$, showing stochastization onset at $I_{\text{I-coil}} \approx$ 1kA. The slope of the curve increases as more and more laminar surface break. The last laminar surface (barrier) breaks at $I_{\text{I-coil}} \approx$ 1.575kA.

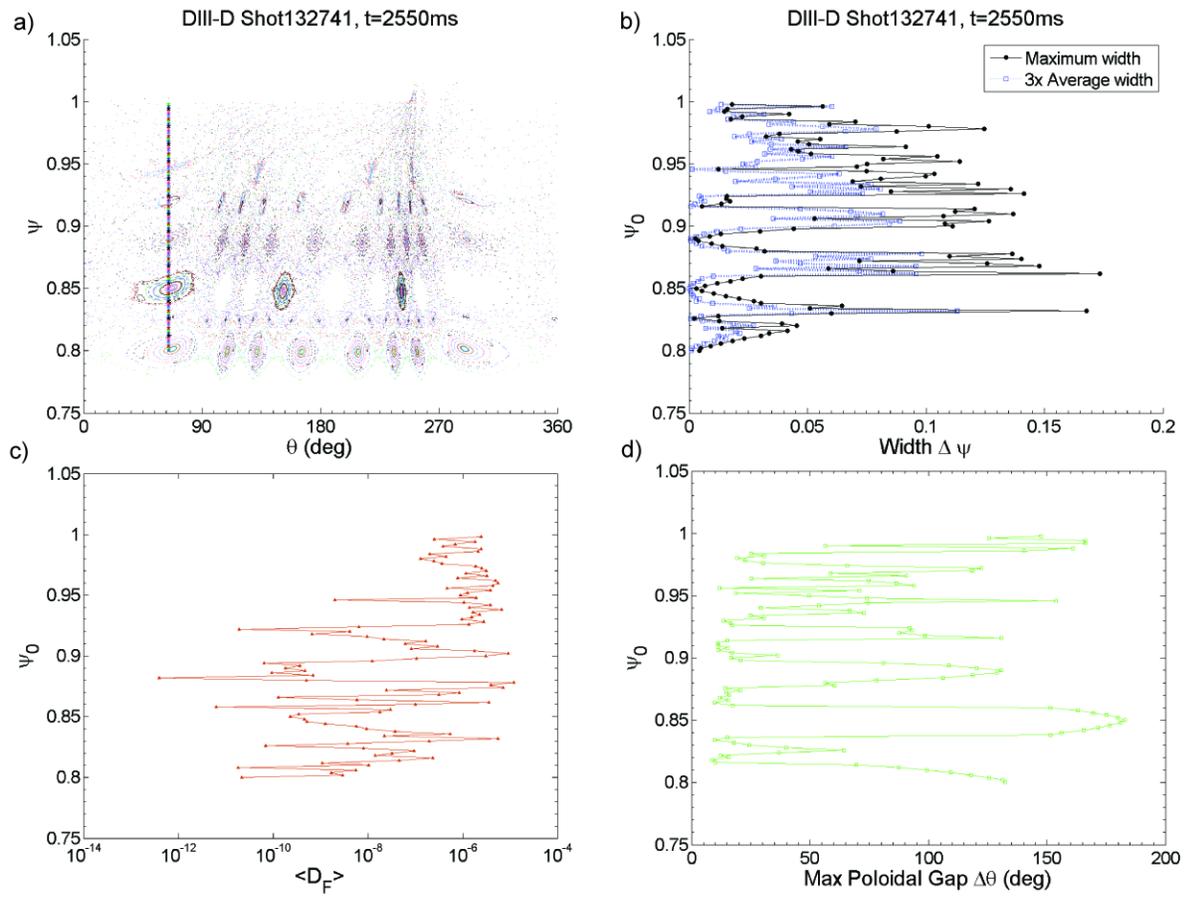

Fig.11 (a) Poincaré plot and corresponding radial profiles of (b) maximum structure width, (c) field-line dispersion and (d) maximum poloidal gap. Symbols at $\theta=65^{o}$ in Fig.(a) denote the initial coordinates of the field-lines traced.

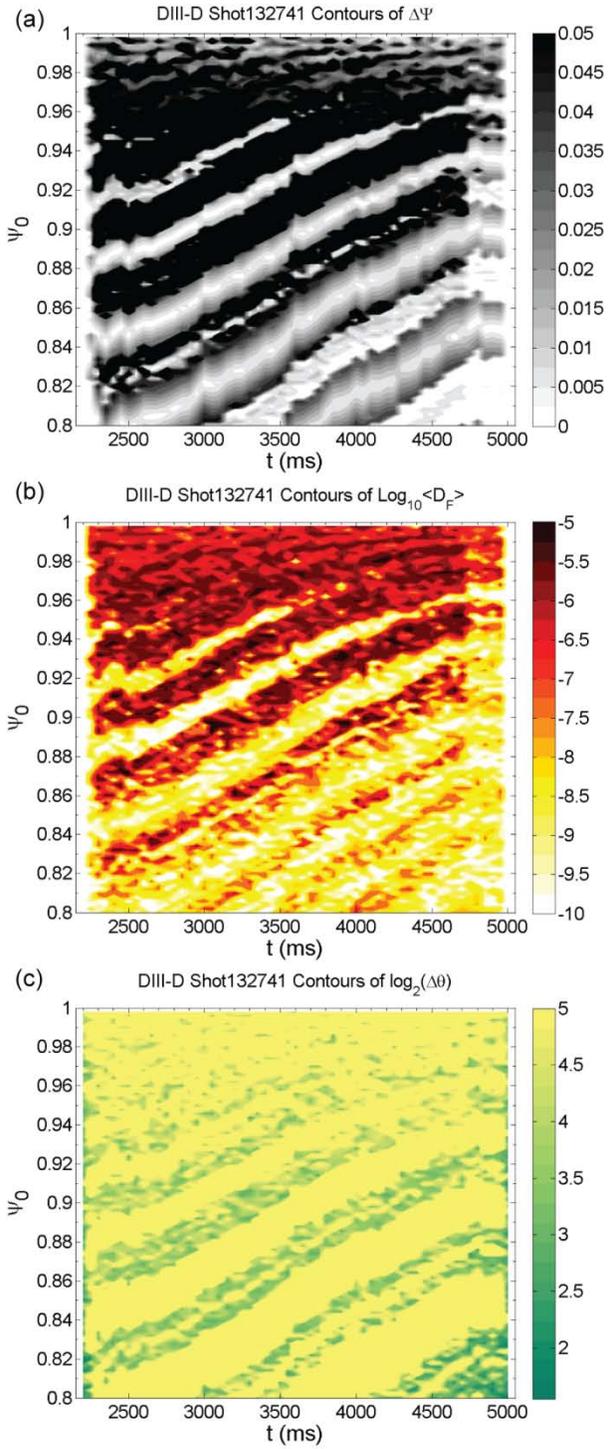

Fig.12 Contours of "structure" (a) width, (b) field-line dispersion and (c) maximum poloidal gap as functions of the radial coordinate (toroidal flux $\Psi$) and time. Zeros and minima denote candidate magnetic barriers and Cantori.

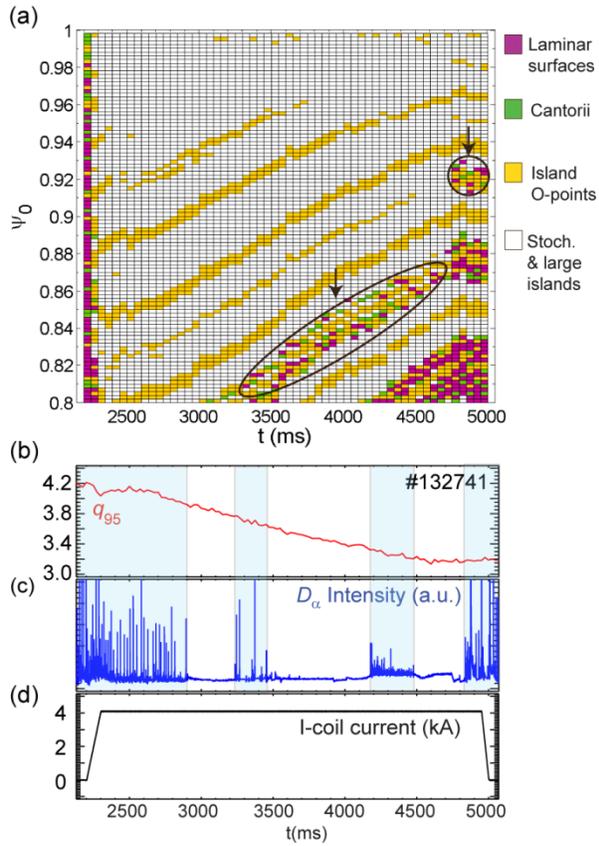

Fig.13 (a) Radial locations of various structures as a function of time. (b-d) Time traces of $q_{95}$, $D_\alpha$ and current in the I-coils used to apply the resonant magnetic perturbation. The outermost laminar surfaces and cantori are marked in Fig.a. Time intervals with no ELM suppression are shaded in Figs.b-c.

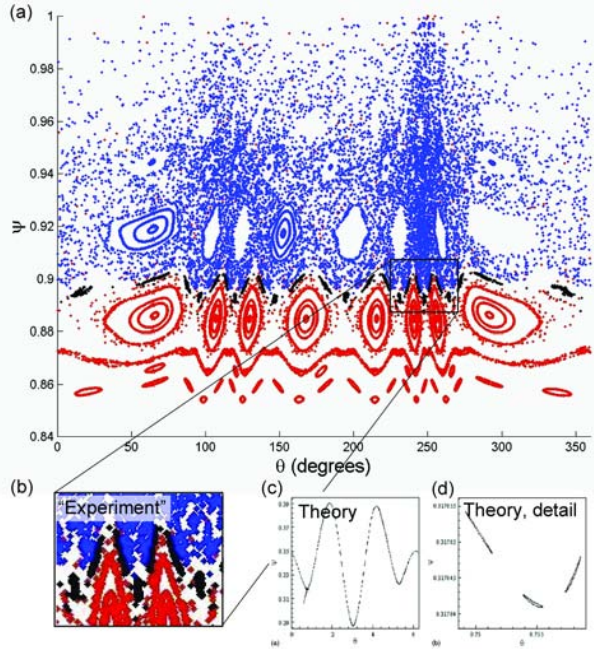

Fig.14 (a) Poincaré plot from DIII-D discharge 136240, exhibiting (b) Cantori similar to (c-d) theory (adapted from Ref.[21]). Color-coding in (a) illustrates that very few field-lines originating below the Cantor set access the region above it, and vice versa, showing that the Cantor set acts as a semi-permeable barrier.

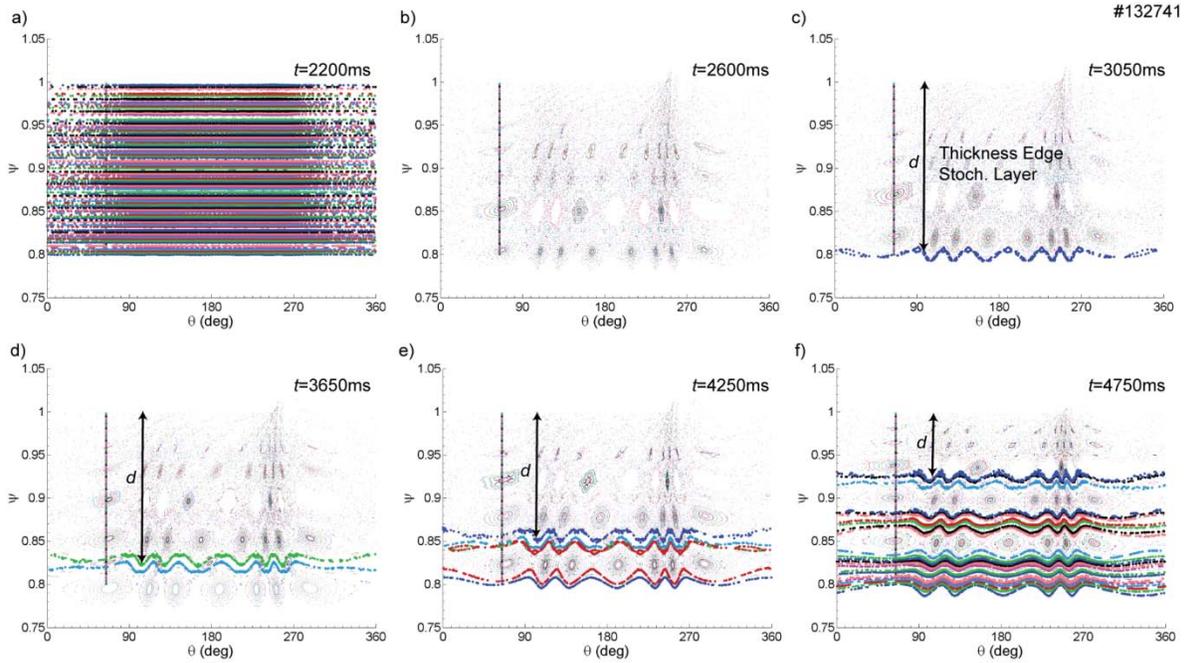

Fig.15 Poincaré plots at various times during a $q_{95}$ ramp-down discharge, showing the formation of magnetic barriers and Cantori (marked by larger symbols), their drift towards outer radii and effect on the thickness $d$ of the edge stochastic layer. Symbols at $\theta=65^o$ denote the initial coordinates of the field-lines traced.

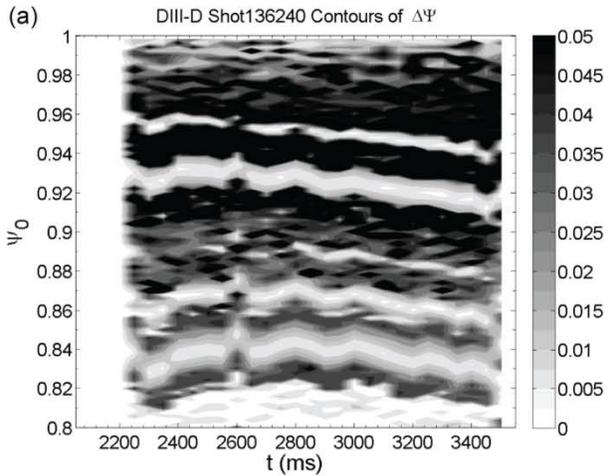

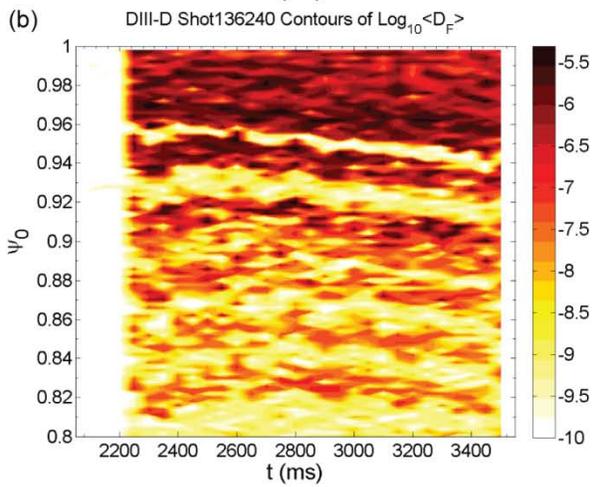

Fig.16 Like Fig.12, but for $q_{95}$ ramp-up discharge 136240.

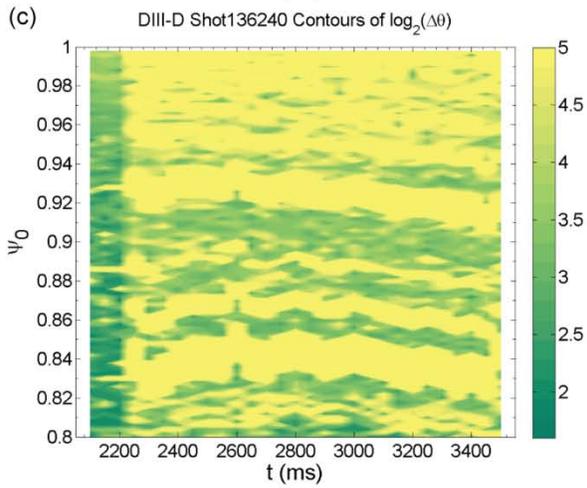

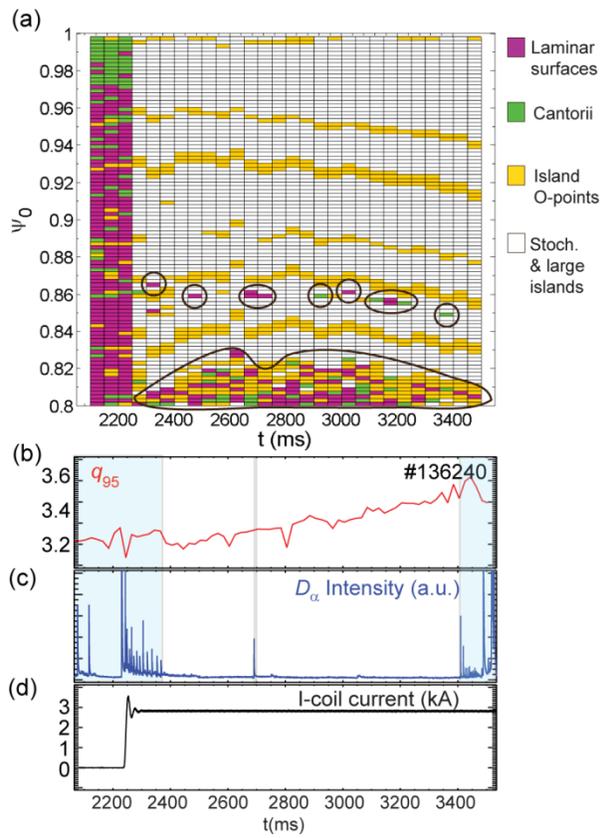

Fig.17 Similar to Fig.13, but for $q_{95}$ ramp-up discharge 136240. Intermittent laminar surfaces and cantorii at $\psi_0 \simeq 0.86$ and persistent ones at $\psi_0 \simeq 0.8$-$0.82$ are marked.

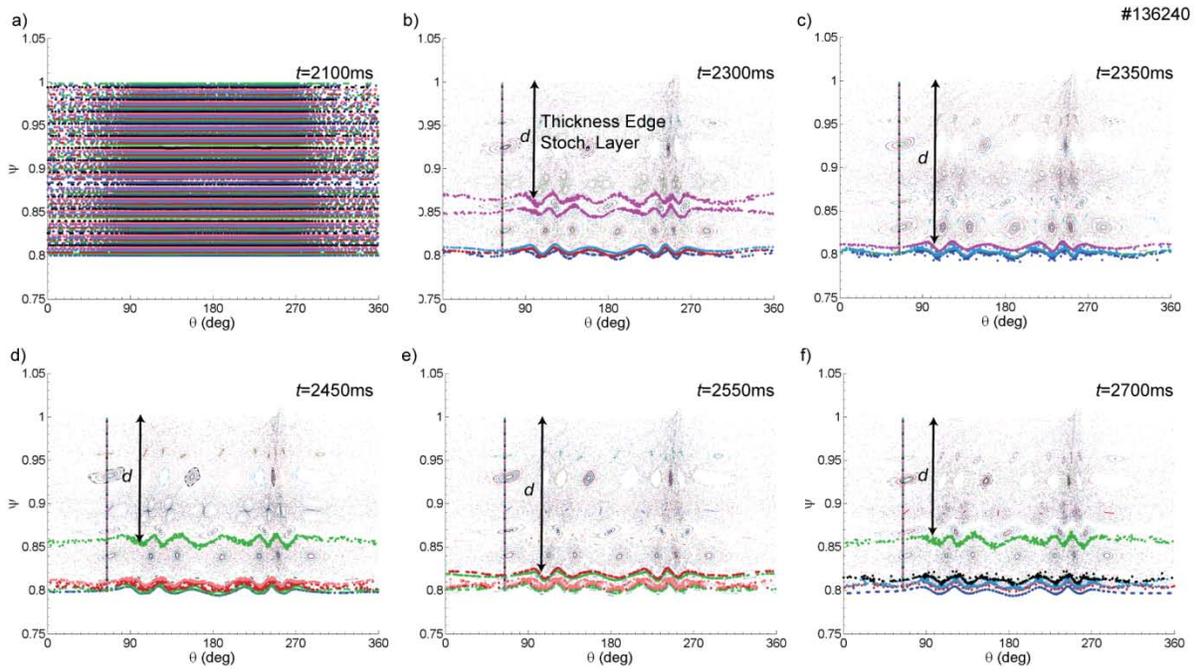

Fig.18 Like Fig.15, but for $q_{95}$ ramp-up discharge 136240.